\documentclass[a4paper,12pt]{article}
\usepackage[latin1]{inputenc}
\usepackage{graphicx}
\usepackage{xcolor}
\usepackage{amsmath}
\usepackage{amsfonts}
\usepackage{cite}

\usepackage{ulem}

\newtheorem{thm}{Theorem}

\newtheorem{la}[thm]{Lemma}

\newtheorem{cor}[thm]{Corollary}
\newtheorem{prop}[thm]{Proposition}

\newcommand{\white}{\qquad{\framebox{\rule{0pt}{4pt}}}}
\newcommand{\owari}{\hfill\white}

\begin{document}
\begin{center}
{\Large Pebble Exchange Group of Graphs}
\end{center}
\begin{center}
{\large Tatsuoki Kato
\footnote{
Email: {\texttt{tatsuoki-k@jm.kj.yamagata-u.ac.jp}}}
}
\end{center}
\begin{center}
{
Yamagata University \\%
2-2-2 Iida-Nishi, Yamagata 990-9585, Japan
}
\end{center}
\begin{center}
{\large Tomoki Nakamigawa
\footnote{This work was supported by JSPS KAKENHI Grant Number JP16K05260.\\
Email: {\texttt{nakami@info.shonan-it.ac.jp}}}
}
\end{center}
\begin{center}
{
Department of Information Science \\%
Shonan Institute of Technology \\%
1-1-25 Tsujido-Nishikaigan, Fujisawa 251-8511, Japan
}
\end{center}
\begin{center}
{\large Tadashi Sakuma
\footnote{This work was supported by JSPS KAKENHI Grant Number JP16K05260, JP26400185, JP18K03388.\\
Email: {\texttt{sakuma@sci.kj.yamagata-u.ac.jp}}}
}
\end{center}
\begin{center}
{
Faculty of Science \\%
Yamagata University \\%
1-4-12 Kojirakawa, Yamagata 990-8560, Japan
}
\end{center}
\begin{abstract}
A graph puzzle ${\rm Puz}(G)$ of a graph $G$ is defined as follows.
A configuration of ${\rm Puz}(G)$ is a bijection from the set of vertices of a board graph to the set of vertices of a pebble graph, both graphs being isomorphic to some input graph $G$. 
A move of pebbles is defined as exchanging two pebbles which are adjacent on both a board graph and a pebble graph.
For a pair of configurations $f$ and $g$, we say that $f$ is equivalent to $g$ if $f$ can be transformed into $g$
by a finite sequence of moves. 

Let ${\rm Aut}(G)$ be the automorphism group of $G$, and let ${\rm 1}_G$ be the unit element of ${\rm Aut}(G)$.
The pebble exchange group of $G$, denoted by ${\rm Peb}(G)$, is defined as the set of all automorphisms $f$ of $G$ such that ${\rm 1}_G$ and $f$ are equivalent to each other.

In this paper, some basic properties of ${\rm Peb}(G)$ are studied.
Among other results, it is shown that for any connected graph $G$, all automorphisms of $G$ are contained in ${\rm Peb}(G^2)$, where $G^2$ is a square graph of $G$.
\end{abstract}
keywords: pebble motion, motion planning, graph puzzle, automorphism

\section{Introduction}
Let $G$ be a finite and undirected graph with no multiple edge or loop.
The vertex set of $G$ and the edge set of $G$ are denoted by $V(G)$ and $E(G)$,
respectively. 
Let $P=\{1,\ldots,k\}$ be a set of pebbles with $k < |V(G)|$.
An \textit{arrangement} of $P$ on $G$ is defined as a function $f$ from $V(G)$
to $\{0, 1, \ldots, k \}$ with $|f^{-1}(i)| = 1$ for $1 \le i \le k$,
where $f^{-1}(i)$ is a vertex occupied with the $i$th pebble for $1 \le i \le k$
and $f^{-1}(0)$ is a set of unoccupied vertices.
A \textit{move} is defined as shifting a pebble from a vertex to
some unoccupied neighbour. 
The {\it pebble motion problem on the pair $(G,P)$} is
to decide whether a given arrangement of pebbles reachable from another
by executing a sequence of moves.
The well-known puzzle named ``15-puzzle'' due to Loyd~\cite{L1959} is
a typical example of this problem where the graph $G$ is a $4 \times 4$-grid.

The pebble motion problem is studied intensively~\cite{A1999,AMPP1999,CDP2006, 
FG2009,FNS2012,J1879,KMS1984,PRST1994,RW1990,S1879,W1974},
because of its considerable theoretical interest as well as
its wide range of applications for computer science and robotics,
such as management of indivisible packets of data moving on
wide-area communication network and motion planning of
independent robots.
In 1974, Wilson\cite{W1974} solved completely 
the feasibility problem (i.e.~the problem of determining 
whether all the configurations of the puzzle are rearrangeable 
from one another or not) for the case of $|f^{-1}(0)| = 1$
on general graphs, and it followed by the result of 
Kornhauser, Miller and Spirakis (FOCS '84)\cite{KMS1984}
for the case of $|f^{-1}(0)| \geq 2$.
Papadimitriou, Raghavan, Sudan and Tamaki
(FOCS '94)\cite{PRST1994} consider the case that
there exists a single special pebble (``robot'') and that 
the other pebbles (``obstacles'') are indistinguishable. 
They focus on the time complexity problems for optimal
number of moves from an arbitrary given arrangement of
the pebbles to a proper goal arrangement in which the
robot is on the desired vertex. 
In 2012, Fujita, Nakamigawa and Sakuma\cite{FNS2012}
generalized the problem to the case of ``colored pebbles'',
where each pebble of $P$ is distinguished by its color.
They also completely solved the feasibility problem for
their model.

In 2015, Fujita, Nakamigawa and Sakuma~\cite{FNS2015}
generalized the pebble motion problem as follows:
For two graphs $G$ and $H$ with a common number of vertices, let us consider
a puzzle ${\rm Puz}(G,H)$, where $G$ is a board graph and $H$ is a pebble graph.
We call a bijection $f$ from $V(G)$ to $V(H)$ a {\it configuration} of ${\rm Puz}(G,H)$,
and we denote the set of all configurations of ${\rm Puz}(G,H)$ by ${\cal C}(G,H)$.
Given a configuration $f$, if $f(x)=y$, we consider that the vertex $x$ of the board
is occupied by the pebble $y$.
In ${\rm Puz}(G,H)$, two pebbles $y_1 = f(x_1)$ and $y_2 = f(x_2)$ can be exchanged
if $x_1 x_2 \in E(G)$ and $y_1 y_2 \in E(H)$.
Then the resultant configuration $g$ satisfies that $g(x_1) = y_2$, $g(x_2) = y_1$ and
$g(x)=f(x)$ for any $x \in V(G) \setminus \{ x_1, x_2 \}$.
We call the operation a {\it move}.
If a configuration $f$ is transformed into another configuration $g$ with a finite
sequence of moves, we say that $f$ and $g$ are {\it equivalent}, denoted by $f \sim g$. 
${\rm Puz}(G,H)$ is called {\it feasible} if all the configurations of the puzzle are
equivalent to each other.

In \cite{FNS2015}, the above mentioned graph puzzle was formally
introduced and some more necessary/sufficient conditions of
the feasibility of the puzzle was studied. 
This model has again a wide range of real world applications, 
especially for robot motion planning problems and facility relocation 
problems.
Please see \cite{FNS2015} for details. 

In this paper, we will shed light on some algebraic property of the
puzzle, which is of not only theoretical interest, but also practical
importance, as will be discussed later.

In the following, we only consider the case where a board graph
and a pebble graph are the same, and we denote ${\cal C}(G,G)$ and
${\rm Puz}(G,G)$ simply by ${\cal C}(G)$ and ${\rm Puz}(G)$, respectively.

The {\it automorphism group} of a graph $G$, denoted by ${\rm Aut}(G)$,
is the group which consists of all bijections $f$ from $V(G)$ to $V(G)$
such that $f(x_1)f(x_2) \in E(G)$ if and only if $x_1 x_2 \in E(G)$.
Let ${\rm 1}_G$, or simply ${\rm 1}$, denote the identity element of ${\rm Aut}(G)$.
Let us introduce the {\it pebble exchange group} of $G$, denoted by ${\rm Peb}(G)$,
as the group which consists of all automorphisms $f$ of $G$ such that ${\rm 1}_G$
and $f$ are equivalent in ${\rm Puz}(G)$.

In application, when a graph $G$ represents some system,
an automorphism $f$ of $G$ corresponds to a rearrangement of possible arrangements of the system.
If $f$ is contained in ${\rm Peb}(G)$, the corresponding rearrangement is realizable with a sequence of local changes step by step.
Hence if we can show the equation
${\rm Peb}(G) ={\rm Aut}(G)$ here, it means 
that practically all the necessary and sufficient 
arrangements are mutually reachable from one another.

However, it seems to be a highly nontrivial
and difficult problem to characterize completely
the graphs whose pebble exchange groups are
equal to their automorphism groups.
Hence, before to attack this problem directly, 
in this paper we will show that the class of
graphs $G$ satisfying ${\rm Peb}(G) ={\rm Aut}(G)$ 
is considerably large.
Especially, we prove (Theorem 12) that,
for any connected graph $G$, the pebble exchange group
of the square $G^2$ of $G$ contains a subgroup isomorphic
to the automorphism group of $G$.
Since the maximum degree of $G^2$ is no more than 
the square of the maximum degree of $G$, 
if the maximum degree of $G$ is a small constant and 
the order of $G$ is sufficiently large, then
not only $G$ but also $G^2$ are sparse graphs, 
and the puzzle ${\rm Puz}(G^2)$
is also far from feasible in general. 
In spite of this, somewhat surprisingly, 
by using Theorem 12, for example, we can show
that, for any connected graph $G$, if we 2-subdivide 
all the edges of $G$, and if we take its square, 
the resulting graph $H$ satisfies the equation
${\rm Peb}(H) ={\rm Aut}(H)$.

\section{Preliminaries}

Pebble motion problems on graphs have been extensively studied.
In the following, let us introduce previously proven theorems closely related to this paper,
by using the concept of ${\rm Puz}(G,H)$.

For two graphs $G$ and $H$, let $G \times H$ denote a
{\it Cartesian product} of $G$ and $H$, where $V(G \times H) = V(G) \times V(H)$ and 
$E(G \times H) = \{ (u_1,v_1)(u_2,v_2) \in V(G \times H)^2 \,:\, u_1 u_2 \in E(G)
{\rm ~and~} v_1=v_2, {\rm ~or~} u_1= u_2 {\rm ~and~} v_1 v_2 \in E(H) \}$.
Let $P_k$ be the path with $k$ vertices, and let $K_{1,\ell}$ be the star with $\ell$ pendant
vertices.

\begin{thm}[W. W. Johnson\cite{J1879}, W. E. Story\cite{S1879}] 
${\rm Puz}(P_4 \times P_4, K_{1,15})$ corresponds to the {\rm 15}-puzzle, by considering that the center $z$
 of $K_{1,15}$ corresponds to the unoccupied space of the {\rm 15}-puzzle. 
For two configurations $f, g \in {\cal C}(P_4 \times P_4, K_{1,15})$ 
with $f^{-1}(z) = g^{-1}(z)$,
$f$ is equivalent to $g$ if and only if $g^{-1} \circ f$ is an even permutation on $V(G)$. 
\end{thm}

We will show some more examples, from Theorem 2 to Theorem 5,
which are considered as generalizations of Theorem 1.   
Suppose that both $G$ and $H$ are bipartite graphs with at least three vertices.
It is not difficult to see that ${\rm Puz}(G,H)$ is not feasible because of the parity of configurations (cf. \cite{FNS2015}).

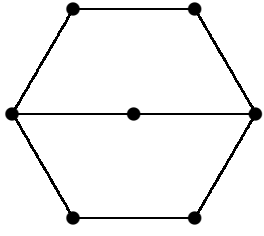
\begin{figure}
\centering
\setlength{\unitlength}{1.6mm}
\begin{center}
\begin{picture}(20,23)(-10,-12)
\put(10,0){\circle*{1}}
\put(-10,0){\circle*{1}}
\put(5,8.66){\circle*{1}}
\put(5,-8.66){\circle*{1}}
\put(-5,8.66){\circle*{1}}
\put(-5,-8.66){\circle*{1}}
\put(0,0){\circle*{1}}
\thinlines
\qbezier(10,0)(7.5,4.33)(5,8.66)
\qbezier(-10,0)(-7.5,-4.33)(-5,-8.66)
\qbezier(5,8.66)(0,8.66)(-5,8.66)
\qbezier(-5,-8.66)(0,-8.66)(5,-8.66)
\qbezier(10,0)(7.5,-4.33)(5,-8.66)
\qbezier(-10,0)(-7.5,4.33)(-5,8.66)
\qbezier(4,0)(7,0)(10,0)
\qbezier(-4,0)(0,0)(4,0)
\qbezier(-4,0)(-7,0)(-10,0)
\end{picture}
\end{center}
\caption{The graph $\theta(1,2,2)$}
\label{fig:theta122}
\end{figure}

Let $\theta(1,2,2)$ be a graph such that $V(\theta(1,2,2))$
$=\{ v_i \,:\, 1\le i\le 7 \}$ and $E(\theta(1,2,2))$
$=\{ v_1 v_2, v_2 v_3, v_3 v_4, v_4 v_5, v_5 v_6, v_6 v_1, v_1 v_7, v_4 v_7 \}$. 
See Figure~\ref{fig:theta122}. 

\begin{thm}[R. M. Wilson\cite{W1974}]
Let $G$ be a $2$-connected non-bipartite graph with $n$ vertices.
If $G$ is not a cycle or $\theta(1,2,2)$, then ${\rm Puz}(G, K_{1,n-1})$ is feasible. 
\end{thm}

For a positive integer $k$, 
a path $P=v_1 v_2 \cdots v_k$ of a graph $G$ is called a $k$-{\it isthmus} if 
(1) every edge of $P$ is a bridge of $G$, 
(2) every vertex of $P$ is a cut-vertex of $G$, and
(3) ${\rm deg}_G (v_i)=2$ for $1 < i < k$.
For two graphs $G$ and $H$, the {\it join} $G+H$ is defined as $V(G+H)=V(G) \cup V(H)$, $E(G+H)=E(G) \cup E(H) \cup \{ uv\,:\,u\in V(G), v\in V(H) \}$.

We remark that an isthmus is intuitively ``a path connecting two blocks'', and we cannot drop any one of conditions (1), (2) and (3) in the definition.
For example, let $G$ be a graph such that $V(G) = \{ v_i \,:\, 1 \le i \le 6 \}$, $E(G) = \{ v_i v_{i+1} \,:\, 1 \le i \le 4 \} \cup \{ v_3 v_6 \}$ and let $P = v_2 v_3 v_4$.
Then $P$ satisfies (1) and (2), but $P$ is not called an isthmus.
Next, let $G$ be a graph such that $V(G) = \{ v_i \,:\, 1 \le i \le 5 \}$, $E(G) = \{ v_i v_{i+1} \,:\, 1 \le i \le 4 \} \cup \{ v_2 v_4 \}$ and let $P = v_2 v_4$.
Then $P$ satisfies (2) and (3), but $P$ is not called an isthmus. 
Finally, let $G$ be a path of $3$ vertices and let $P=G$.
Then $P$ satisfies (3) and (1), but $P$ is not called an isthmus.    

Let $K_n$ denote a complete graph with $n$ vertices, 
and let $K_{m,n}$ denote a complete bipartite graph on $m$ and $n$ vertices. 
For a graph $G$, let $\overline{G}$ be the {\it complement} of $G$. 
A set of pebbles is called {\it labeled} if we can distinguish them from one another.
 Otherwise we call the set of pebbles {\it unlabeled}. 

\begin{thm}[D. Kohnhauser, G. Miller, and P. Spirakis\cite{KMS1984}]
Let $2 \le k \le n$.
A pebble graph $K_k + \overline{K_{n-k}}$ is considered as a set of $n-k$ labeled pebbles and  $k$ unlabeled pebbles, in which two labeled pebbles cannot directly exchange their positions with each other.
Let $G$ be a connected graph with $n$ vertices except a cycle.
Then ${\rm Puz}(G, K_k + \overline{K_{n-k}})$ is feasible if and only if $G$ has no $k$-isthmus. 
\end{thm}

\begin{thm}[S. Fujita, T. Nakamigawa, and T. Sakuma\cite{FNS2012}]
Let $2 \le k \le n/2$.
Let $G$ be a graph with $n$ vertices.
Then ${\rm Puz}(G, K_{k,n-k})$ is feasible if and only if 
(1) $G$ is not a cycle, and
(2) $G$ is not bipartite, and
(3) $G$ has no $k$-isthmus. 
\end{thm}

\begin{thm}[S. Fujita, T. Nakamigawa, and T. Sakuma\cite{FNS2012}]
Let $r \ge 3$ and let $2 \le n_1 \le \ldots \le n_r$.
Let $G$ be a graph with $n = n_1 + n_2 + \cdots + n_r$ vertices.
Then ${\rm Puz}(G, K_{n_1, n_2, \ldots, n_r})$ is feasible if and only if 
(1) $G$ is not a cycle, and
(2) $G$ has no $(n-n_r)$-isthmus. 
\end{thm}

Now we show that ${\rm Peb}(G)$ is a group,
more precisely, a normal subgroup of ${\rm Aut}(G)$, first
we prepare the following lemma.
\begin{la}\label{peb group lemma}
For $\varphi_1$, $\varphi_2$, $\alpha$, $\beta$ $\in {\rm Aut}(G)$, 
if $\varphi_1 \sim \varphi_2$, then we have $\beta \circ \varphi_1 \circ \alpha \sim \beta \circ \varphi_2 \circ \alpha $.
\end{la}
{\bf Proof. }
Since $\varphi_1 \sim \varphi_2$, there exists a finite sequence of
transpositions $\sigma_1, \sigma_2, \ldots, \sigma_s$ on $V(G)$
corresponding to moves from $f_0 = \varphi_1$ to $f_s = \varphi_2$ such that
$f_i = \sigma_i \circ f_{i-1}$ for $1 \le i \le s$.
By the definition of a move in ${\rm Puz}(G)$, for $1 \le i \le s$,
we have a pair of vertices $x_i$ and $y_i$ satisfying
$(x_i, y_i) \in E(G)$, $(f_{i-1} (x_i), f_{i-1} (y_i)) \in E(G)$ and
$\sigma_i$ exchanges $f_{i-1}(x_i)$ and $f_{i-1}(y_i)$.

Put $\varphi'_j = \beta \circ \varphi_j \circ \alpha$ for $1 \le j \le 2$.
Let us define a sequence of configurations $g_i = \beta \circ f_i \circ \alpha$ for $0 \le i \le s$ and let us define a sequence of transpositions $\tau_i = \beta \circ \sigma_i \circ \beta^{-1}$ for $1\le i \le s$.
Furthermore, let us define $u_i = \alpha^{-1}(x_i)$ and $v_i = \alpha^{-1}(y_i)$.
Then we have $g_i = \beta \circ (\sigma_i \circ f_{i-1}) \circ \alpha = (\beta \circ \sigma_i \circ \beta^{-1}) \circ ( \beta \circ f_{i-1} \circ \alpha) = \tau_i \circ g_{i-1}$ for $1 \le i \le s$.  

What remains to be necessary to check is that $g_i$'s are corresponding to moves.
We have $(u_i, v_i) \in E(G)$ and $(g_{i-1}(u_i), g_{i-1}(v_i)) = (\beta \circ f_{i-1} \circ \alpha(u_i), \beta \circ f_{i-1} \circ \alpha(v_i))  = (\beta \circ f_{i-1} (x_i), \beta \circ f_{i-1} (y_i))  \in E(G)$.
Finally, $\tau_i$ is a transposition exchanging $\beta ( f_{i-1} (x_i))$ and $\beta ( f_{i-1} (y_i))$, namely $g_{i-1} (u_i)$ and $g_{i-1} (v_i)$, as required. 
\owari

\begin{prop}\label{peb normal subgroup}
${\rm Peb}(G)$ is a normal subgroup of ${\rm Aut}(G)$.
\end{prop}
{\bf Proof. }
Firstly, let $f, g \in {\rm Peb}(G)$.
Since ${\rm 1}_G \sim g$, by Lemma \ref{peb group lemma}, we have
$f = {\rm 1}_G \circ {\rm 1}_G \circ f \sim {\rm 1}_G \circ g \circ f = g \circ f$.
Hence, we have ${\rm 1}_G \sim f \sim g \circ f$.
It follows that $g \circ f \in {\rm Peb}(G)$. 
Secondly, By definition, ${\rm 1}_G$ is contained in ${\rm Peb}(G)$.
Therefore, ${\rm Peb}(G)$ is a subgroup of ${\rm Aut}(G)$.
In order to show that ${\rm Peb}(G)$ is a normal subgroup of ${\rm Aut}(G)$,
 what we need to show is that for 
$f \in {\rm Peb}(G)$ and $g \in {\rm Aut}(G)$,
 we have $g \circ f \circ g^{-1} \in {\rm Peb}(G)$.
Since ${\rm 1}_G \sim f$, by Lemma \ref{peb group lemma}, we have ${\rm 1}_G = g \circ {\rm 1}_G \circ g^{-1} \sim g \circ f \circ g^{-1}$.
Hence, we have $g \circ f \circ g^{-1} \in {\rm Peb}(G)$, as required.
\owari 

\section{Main Results}
It is known that for any finite group $\Gamma$, there exists a graph $G$ such that ${\rm Aut}(G) \simeq \Gamma$ (cf. \cite{F1949}).
By using this fact, we have the following result.   

\begin{prop}\label{existence-peb}
For any finite group $\Gamma$, there exists a graph $G$ such that ${\rm Peb}(G) \simeq \Gamma$.  
\end{prop}
{\bf Proof. }
Let us take a graph $H$ such that ${\rm Aut}(H) \simeq \Gamma$.
Since at least one of $H$ and $\overline{H}$ is connected, and ${\rm Aut}(H) \simeq {\rm Aut}(\overline{H})$, 
by replacing $H$ with $\overline{H}$, if necessary,
we may assume $H$ is connected.

Let us build a new graph $H'$ from $H$ as follows;
 $V(H') = V(H) \cup \{ x^i_{uv}\,:\, uv \in E(H), 1 \le i \le 3 \}$,
 $E(H') = \{ ux^1_{uv}, vx^1_{uv}, x^1_{uv}x^2_{uv}, x^2_{uv}x^3_{uv} \,:\, uv \in E(H) \}$.
Intuitively, $H^{\prime}$ is obtained by adding a pendant path to the middle of each edge of $H$. 

{\bf Claim $1$. }
${\rm Aut}(H') \simeq{\rm Aut}(H)$. 
\medskip\\
For $f \in {\rm Aut}(H)$,
 let us define a bijection $f'$ on $V(H')$ such that $f'(v) = f(v)$ for $v \in V(H)$ and $f'(x^i_{uv}) = x^i_{f(u)f(v)}$.
Then we have $f' \in {\rm Aut}(H')$, and by this correspondence, we have ${\rm Aut}(H) \subset {\rm Aut}(H')$.

Conversely, we will show that ${\rm Aut}(H') \subset {\rm Aut}(H)$.
Let $g \in {\rm Aut}(H')$.
For $uv \in V(H)$, $x^1_{uv} x^2_{uv} x^3_{uv}$, denoted by $P_{uv}$ is a pendant path of length two contained in $H'$.
Since $H'$ contains no pendant path of length two other than $P_{uv}$ for some $uv \in E(H)$, 
 we have $g(P_{uv}) = P_{g(u)g(v)}$. 
It follows that an automorphism $g$ of $H'$ restricted on $V(H)$ induces an automorphism of $H$.
Therefore, we have ${\rm Aut}(H') \subset {\rm Aut}(H)$.

Now, let us consider $G=H'+z$ with $n$ vertices, which is the join of $H'$ and an additional vertex $z$.
\medskip\\
{\bf Claim $2$. }
${\rm Aut}(G) \simeq {\rm Aut}(H')$. 
\medskip\\
For $f \in {\rm Aut}(H')$, let $f'$ be a bijection on $V(G)$ such that $f'(z) = z$ and $f'(v) = f(v)$ for $v \in V(H')$.
Then we have $f' \in {\rm Aut}(G)$, and by this correspondence, we have ${\rm Aut}(H') \subset {\rm Aut}(G)$.
Conversely, for $g \in {\rm Aut}(G)$, we have $g(z)=z$, because $z$ is the unique vertex of degree $|V(G)|-1$ in $G$.
Hence, $g$ restricted on $V(H')$ induces an automorphism of $H'$.
Hence, we have ${\rm Aut}(G) \subset {\rm Aut}(H')$, as required.

Since $G$ is a non-bipartite $2$-connected graph containing $K_{1,n-1}$
as a spanning subgraph, by Theorem 2, ${\rm Puz}(G)$ is feasible. 
Therefore, we have ${\rm Peb}(G) = {\rm Aut}(G)$.
Since ${\rm Aut}(G) \simeq {\rm Aut}(H') \simeq {\rm Aut}(H) \simeq \Gamma$, 
we have ${\rm Peb}(G) \simeq \Gamma$, as required. 
\owari

Second, we note a simple observation about ${\rm Peb}(G)$, where $G$ contains no small cycle.
For a graph $G$, the {\it girth} of $G$, denoted by ${\rm girth}(G)$, is the order of a smallest cycle contained in $G$.
If $G$ contains no cycle, ${\rm girth}(G)$ is defined as $\infty$.
A {\it matching} of a graph $G$ is a set of independent edges of $G$. 
For a matching $M$ of a graph $G$, let $f(M)$ be a configuration of ${\rm Puz}(G)$ such that for all $x \in V(M)$, $f(x)=y$, where $xy \in E(M)$, and $f(x)=x$ for all $x \not\in V(M)$.
Let ${\cal M}(G) = \{ f(M) \,:\, M {\rm ~is~a~matching~of~} G \}$. 

\begin{prop}\label{large-girth}
Let $G$ be a connected graph with at least three vertices.
If ${\rm girth}(G) \ge 5$, then ${\rm Peb}(G) \simeq \{ {\rm 1}_G \}$.
\end{prop}
{\bf Proof. } 
It is sufficient to show that if $f \in {\cal C}(G) \setminus \{ {\rm 1}_G \}$ satisfies ${\rm 1}_G \sim f$, then $f$ is not an automorphism of $G$.
\medskip\\
{\bf Claim.} %
Let $f$ be a configuration of $G$.
Then $f \sim {\rm 1}_G$ if and only if $f \in {\cal M}(G)$.
\medskip\\
First, suppose that $f=f(M) \in {\cal M}(G)$, where $M$ is a matching of $G$.
Starting from ${\rm 1}_G$, by exchanging all pairs of pebbles $u$ and $v$ satisfying $uv \in E(M)$, we have $f(M) \sim {\rm 1}_G$.

Second, suppose that $f \sim {\rm 1}_G$.
Let $f_0={\rm 1}_G, f_1, f_2, \ldots, f_{s-1}, f_s=f$ be a sequence of configurations, where $f_{i}$ is generated from $f_{i-1}$ by a move for all $1 \le i \le s$.
By induction, we may assume $f_{s-1}=f(M) \in {\cal M}(G)$.
Let us assume we have $f$ from $f_{s-1}$ by a move, in which two pebbles $u$ and $v$ are exchanged. 
What we need to show is that $f \in {\cal M}(G)$.
\medskip\\
\underline{Case 1.} Both $u$ and $v$ are contained in $V(M)$ and $uv \in E(M)$.
\medskip\\
In this case, we have $f=f(M')$, where $M'=M \setminus \{ uv \}$.
\medskip\\
\underline{Case 2.} Both $u$ and $v$ are contained in $V(M)$ and $uv \not\in E(M)$.
\medskip\\
Suppose that $ux \in E(M)$ and $vy \in E(M)$.
In order to exchange $u=f_{s-1}(x)$ and $v=f_{s-1}(y)$, we have $uv \in E(G)$ and $xy \in E(G)$.
Hence, $uxyv$ forms a cycle of length $4$, a contradiction.
\medskip\\
\underline{Case 3.} Exactly one of $u$ and $v$ is contained in $V(M)$.
\medskip\\
We may assume $u \in V(M)$ and $v \not\in V(M)$.
Suppose that $ux \in E(M)$.
In order to exchange $u=f_{s-1}(x)$ and $v=f_{s-1}(v)$, we have $uv \in E(G)$ and $vx \in E(G)$.
Hence, $uxv$ forms a cycle of length $3$, a contradiction.
\medskip\\
\underline{Case 4.} 
Neither $u$ nor $v$ is contained in $V(M)$.  
\medskip\\
In this case, we have $f=f(M')$, where $M'=M \cup \{ uv \}$.
\medskip
Suppose, for contradiction, that there exists an automorphism $f$ of $G$ 
with $f \sim {\rm 1}_G$ and $f \ne {\rm 1}_G$. 
By the above claim, we have a matching $M$ of $G$ such that $f=f(M)$.
Since $f \ne {\rm 1}_G$, we have $E(M) \ne \emptyset$.
Let $uv \in E(M)$.
Because $|V(G)|$ is at least $3$ and $G$ is connected, we may assume there exists a vertex $x \in V(G) \setminus \{ u,v \}$ such that $ux \in E(G)$.
If $x \in V(M)$, there exists an edge $xy \in E(M)$.
Since $f$ is an automorphism of $G$, we have $f(u)f(x) = vy \in E(G)$.
Hence, $uvyx$ forms a cycle of length $4$, a contradiction.
If $x \not\in V(M)$, since $f$ is an automorphism of $G$, we have $f(u)f(x) = vx \in E(G)$.
Hence, $uvx$ forms a cycle of length $3$, a contradiction.
\owari
\medskip

The next result is about the pebble exchange group of a product of graphs.

For two graphs $G_1$ and $G_2$,  
the {\it Cartesian product} of $G_1$ and $G_2$, denoted by $G_1 \times G_2$, is a graph such that 
its vertices are ordered pairs of elements $(x_1, x_2)$, where $x_i \in V(G_i)$ for $1 \le i \le 2$,
and two vertices $(x_1, x_2)$ and $(y_1, y_2)$ of $G_1 \times G_2$ are adjacent if and
only if either $x_1 = y_1$ and $x_2 y_2$ is an edge of $G_2$ or $x_2 = y_2$ and $x_1 y_1$ is an edge
of $G_1$. 
For two groups $\Gamma_1$ and $\Gamma_2$,
the {\it direct product} of $\Gamma_1$ and $\Gamma_2$, denoted by $\Gamma_1 \times \Gamma_2$,
is a group such that its elements are ordered pairs of elements $(\alpha_1, \alpha_2)$,
where $\alpha_i \in \Gamma_i$ for $1 \le i \le 2$, and its multiplication $\circ$ is defined as
$(\alpha_1, \alpha_2) \circ (\beta_1, \beta_2) = (\alpha_1 \circ_1 \beta_1, \alpha_2 \circ_2 \beta_2)$,
where $\circ_i$ is the multiplication of $\Gamma_i$ for $1 \le i \le 2$.

\begin{thm}\label{graph-product}
For any two connected graphs $G_1$ and $G_2$, ${\rm Peb}(G_1 \times G_2) \simeq {\rm Peb}(G_1) \times {\rm Peb}(G_2)$.  
\end{thm}

The proof of Theorem \ref{graph-product} will be given in Section $3$.

Let $Q_n$ be the $n$-dimensional hypercubic graph. 
Since $Q_n = P_2^n$ and ${\rm Peb}(P_2) \simeq {\mathbb Z}_2$, 
we have the following corollary as an immediate consequence of Theorem \ref{graph-product}.

\begin{cor}\label{hyper-cube}
For $n \ge 1$, ${\rm Peb}(Q_n) \simeq ({\mathbb Z}_2)^n$.
\end{cor}

As a graph $G$ becomes sparse, the number of possible moves on $G$ decreases. 
Hence, it is interesting to show the existence of graphs $G$ such that ${\rm Peb}(G)$ has a rich structure and $|E(G)| = O(|V(G)|)$.

For two vertices $u,v$ of a graph $G$, let $d_G(u,v)$ denote the {\it distance}
between $u$ and $v$ in $G$. Furthermore, for two subsets $X,Y$ of the vertex set
$V(G)$, let us define $d_G(X,Y):=\min\{d_G(u,v) | u \in X, v \in Y\}$.
We abbreviate $d_G(\{u\},Y)$ (resp. $d_G(X,\{v\})$) to $d_G(u,Y)$ (resp. $d_G(X,v)$).

The {\it square graph} $G^2$ of $G$ is defined as
 $V(G^2) = V(G)$ and $E(G^2) = \{ uv \in V(G)^2 \,:\, d_G(u,v)=1 {\rm ~or~} 2  \}$. 

The main result of the paper is the following theorem.

\begin{thm}\label{graph-square}
For any connected graph $G$, ${\rm Peb}(G^2) \supset {\rm Aut}(G)$.
\end{thm}

In order to prove Theorem \ref{graph-square}, we first deal with the simplest but the most important case, where $G$ is a path.
 
\begin{la}\label{path-square}
For $n \ge 2$, ${\rm Peb}(P_n^2) \supset {\rm Aut}(P_n)$.
\end{la}

The proof of Lemma \ref{path-square} will be given in Section $4$.

Second, let us introduce a new operation, {\it path flip}, for a configuration $f \in {\cal C}(G)$.
Let $P = v_0 v_1 \ldots v_n$ be a path of $G$.
If $f(v_0) f(v_1) \ldots f(v_n)$ is also a path of $G$, by a path flip, $f$ can be replaced with  $g \in {\cal C}(G)$ such that $g(v_i)=f(v_{n-i})$ for $0 \le i \le n$, and $g(x)=f(x)$ for all $x \in V(G) \setminus V(P)$.

The following lemma may be of independent interest apart from pebble exchange puzzles.

\begin{la}\label{path-flip}
For a connected graph $G$, and for any two configurations $f$, $g \in {\rm Aut}(G)$, $f$ can be transformed into $g$ by a finite sequence of path flips.
\end{la}

The proof of Lemma \ref{path-flip} will be given in Section $5$.

By Lemma \ref{path-square}, any path flip can be achieved with a sequence of pebble exchanges in ${\rm Puz}(G^2)$.
Hence, by Lemma \ref{path-flip}, Theorem \ref{graph-square} follows.

\section{Proof of Theorem \ref{graph-product}}
First, we will show that ${\rm Peb}(G_1) \times {\rm Peb}(G_2) \subset {\rm Peb}(G_1 \times G_2)$.  
For $\sigma \in {\rm Peb}(G_1)$ and $\tau \in {\rm Peb}(G_2)$, it suffices to show that $(\sigma,\tau) \in {\rm Peb}(G_1 \times G_2)$.
In the first part of moves, we process a sequence of moves corresponding to $\sigma$ on all copies of $G_1$ in parallel.
In the second part of moves, we process a sequence of moves corresponding to $\tau$ on all copies of $G_2$ in parallel.
The sequence of all moves yields $(\sigma,\tau)$.

Second, we will show that ${\rm Peb}(G_1 \times G_2) \subset {\rm Peb}(G_1) \times {\rm Peb}(G_2)$.
\medskip\\
{\bf Claim 1.} %
Let $f \in {\cal C}(G_1 \times G_2)$ such that ${\rm 1}_{G_1 \times G_2} \sim f$. 
For two pebbles $x$ and $y$,
 if $f^{-1}(x)$ and $f^{-1}(y)$ are in a common copy of $G_i$ for some $i=1,2$, then $x$ and $y$ are not in a common copy of $G_{3-i}$.
\medskip\\
Suppose, for contradiction, that there exists a pair of pebbles $x$ and $y$ and a configuration $f$ with ${\rm 1} \sim f$ such that
$f^{-1}(x)$ and $f^{-1}(y)$ are in a common copy of $G_i$, and $x$ and $y$ are in a common copy of $G_{3-i}$.
We may assume that $f$ can be reached from ${\rm 1}$ with the minimum number $s$ of moves
until at least one of such counterexamples $(x,y)$ occurs. 
We may assume that $y$ is exchanged with a pebble $z$ in the $s$-th move.
\medskip\\
\underline{Case 1.} $f^{-1}(y)$ and $f^{-1}(z)$ are in a common copy of $G_{3-i}$.
\medskip\\
In this case, by the minimality of $s$, $y$ and $z$ are in a common copy of $G_{3-i}$.
Since $x$ and $y$ are in a common copy of $G_{3-i}$, $x$ and $z$ are in a common copy of $G_{3-i}$.
Then the pair $(x,z)$ becomes our counterexample 
just after the $(s-1)$-th steps.
This contradicts the minimality of $s$. 
\medskip\\
\underline{Case 2.} $f^{-1}(y)$ and $f^{-1}(z)$ are in a common copy of $G_i$.
\medskip\\
In this case,
the pair $(x,y)$ is already our counterexample 
after the $(s-1)$-th steps.
This contradicts the minimality of $s$. \owari 
\medskip\\
{\bf Claim 2.} %
Let $f$ be an automorphism of $G_1 \times G_2$ such that $1_{G_1 \times G_2} \sim f$.
For two pebbles $x$ and $y$, if $f^{-1}(x)$ and $f^{-1}(y)$ are in a common copy of $G_i$ for some $i=1,2$, then $x$ and $y$ are in a common copy of $G_i$. 
\medskip\\
Let us assume that $f^{-1}(x)$ and $f^{-1}(y)$ are in a common copy of $G_i$.
Since $f$ is an automorphism of $G_1 \times G_2$, there exists a path $P$ from $x$ to $y$ such that a path  $f^{-1}(V(P))$ is in a common copy of $G_i$.
By Claim 1, all pairs of vertices in $V(P)$ are in a mutually different copy of $G_{3-i}$.
Hence, any pair of adjacent vertices in $V(P)$ are in a common copy of $G_i$.
Therefore, $x$ and $y$ are in a common copy of $G_i$. \owari 

By Claim 2, $f$ induces a permutation $\tilde{\sigma}_i$ on the set of all copies of $G_{3-i}$ for $i=1,2$, where $\tilde{\sigma}_i$ naturally corresponds to $\sigma_i \in {\rm Aut}(G_i)$.
Then, we have $f = (\sigma_1,\sigma_2) \in {\rm Aut}(G_1) \times {\rm Aut}(G_2)$.
Furthermore, by Claim 1, if two pebbles $x$ and $y$ are in a common copy of $G_i$,
$x$ and $y$ can be exchanged only if they occupy a common edge of a common copy of $G_i$.
Hence, we have $\sigma_i \in {\rm Peb}(G_i)$ for $i=1,2$.
Therefore, we have $f \in {\rm Peb}(G_1) \times {\rm Peb}(G_2)$.
\owari

\section{Proof of Lemma \ref{path-square}}
It is not difficult to see that ${\rm Puz}(P_n^2)$ is feasible for $n \le 5$.
Hence, in this case, we have  ${\rm Peb}(P_n^2) = {\rm Aut}(P_n^2) \supset {\rm Aut}(P_n)$.
Suppose that $n \ge 6$.
In this case, since ${\rm Aut}(P_n^2) = {\rm Aut}(P_n) \simeq {\mathbb Z}_2$,
it suffices to prove ${\rm Peb}(P_n^2) \simeq {\mathbb Z}_2$. 
Let us label the vertices of $P_n$ as $V(P_n) = \{ 1,2,\ldots,n \}$ and $E(P_n) = \{ i j \,:\, j-i = 1 \}$.
Note that ${\rm Aut}(P_n^2) = \{ {\rm 1}_n, \alpha_n \}$, where ${\rm 1}_n(i)=i$ for $1 \le i \le n$
and $\alpha_n(i)=n-i+1$ for $1 \le i \le n$.
It suffices to show that ${\rm 1}_n \sim \alpha_n$ in ${\rm Puz}(P_n^2)$.

In the following, besides ${\rm Puz}(P_n^2)$,
we consider two additional puzzles 
${\rm Puz}(P_{n+1}^2 \setminus \{ n \}, P_n^2)$ and
${\rm Puz}(P_n^2,P_{n+1}^2 \setminus \{ n \})$.
For configurations
 $f \in {\cal C}(P_n^2,P_n^2)$,
 $g \in {\cal C}(P_{n+1}^2 \setminus \{ n \}, P_n^2)$
 and
 $h \in {\cal C}(P_n^2,P_{n+1}^2 \setminus \{ n \})$,
we will use notations as
\begin{eqnarray*}
f & = & (f(1), f(2), \ldots, f(n-1), f(n)), \\
g & = & (g(1), g(2), \ldots, g(n-1), \ast, g(n+1)), \\
h & = & (h(1), h(2), \ldots, h(n-1), h(n)).
\end{eqnarray*}
By using this notation, ${\rm 1}_n$ and $\alpha_n$ is expressed as 
$$
{\rm 1}_n  =  (1,2,\ldots,n-1,n), \hspace*{1cm}
\alpha_n  =  (n,n-1,\ldots,2,1).
$$
Let us define ${\rm 1}'_n$ and $\beta_n \in {\cal C}(P_{n+1}^2 \setminus \{ n \}, P_n^2)$ as
$$
{\rm 1}'_n  =  (1,2,\ldots,n-1,\ast,n), \hspace*{1cm}
\beta_n  = (n,n-1,\ldots,2,\ast,1),
$$
and let us define ${\rm 1}''_n$ and $\gamma_n \in {\cal C}(P_n^2,P_{n+1}^2 \setminus \{ n \})$ as
$$
{\rm 1}''_n  =  (1,2,\ldots,n-1,n+1), \hspace*{1cm}
\gamma_n  =  (n+1,n-1,\ldots,2,1).
$$
What we want to show is that 
${\rm 1}_n \sim \alpha_n$, 
${\rm 1}'_n \sim \beta_n$, 
${\rm 1}''_n \sim \gamma_n$ for all $n \ge 1$.
\\
Note that $P_{n+1}^2 \setminus \{ n \}$ is naturally considered as a subgraph of $P_n^2$.
Hence, if ${\rm 1}'_n \sim \beta_n$, by using the same sequence of moves from ${\rm 1}'_n$ to $\beta_n$, we have a sequence of moves from ${\rm 1}_n$ to $\alpha_n$. 
Therefore, ${\rm 1}'_n \sim \beta_n$ implies that ${\rm 1}_n \sim \alpha_n$.
Furthermore, ${\rm Puz}(P_{n+1}^2 \setminus \{ n \}, P_n^2)$ and
${\rm Puz}(P_n^2,P_{n+1}^2 \setminus \{ n \})$ are isomorphic as puzzles,
since these puzzles can be switched to each other by interchanging
the roles of a board graph and a pebble graph, and ${\rm 1}'_n$ and $\beta_n$ are
corresponding to ${\rm 1}''_n$ and $\gamma_n$, respectively.
Hence, ${\rm 1}'_n \sim \beta_n$ holds if and only if ${\rm 1}''_n \sim \gamma_n$ holds.

We proceed by induction on $n$.
For $n \le 2$, it is not difficult to see that the conclusion holds.
Let $n \ge 3$.
It suffices to show that ${\rm 1}'_n \sim \beta_n$ by using the inductive assumptions ${\rm 1}_k \sim \alpha_k$, 
${\rm 1}'_k \sim \beta_k$, 
${\rm 1}''_k \sim \gamma_k$ for $2 \le k \le n-1$.
We have
\begin{eqnarray*}
{\rm 1}'_n & = & (1,2,\ldots,n-2,n-1,\ast,n) \\
 & \sim & (1,n,\ldots,4,3,\ast,2) \ \ \ \ {\rm by~}{\rm 1}'_{n-1}\sim \beta_{n-1} \\
 & \sim & (1,3,\ldots,n-1,n,\ast,2) \ \ {\rm by~}{\rm 1}_{n-2}\sim \alpha_{n-2} \\
 & \sim & (n,n-1,\ldots,3,1,\ast,2) \ \ {\rm by~}{\rm 1}''_{n-1}\sim \gamma_{n-1} \\
 & \sim & (n,n-1,\ldots,3,2,\ast,1) \ \ {\rm by~the~exchange~of~}1{\rm ~and~}2 \\
 & = & \beta_n,
\end{eqnarray*}
 as required. 
\owari

\section{Proof of Lemma \ref{path-flip}}
In the following, for a configuration $f \in {\cal C}(G)$, we say that $f$ is {\it realizable} by path flips, if $f$ can be transformed from $1_G$ by a finite sequence of path flips.
Note that, in general, if two automorphisms $\sigma$ and $\tau$ of $G$ are realizable by path flips, their composition $\tau \circ \sigma$ is also realizable by path flips.
(This fact for path flips can be proved in the same way as in the proof of Proposition \ref{peb normal subgroup} for pebble exchanges, which are ``edge flips'', that is, flips of paths of length $1$. )

Suppose, for contradiction, that there exists a pair $(G,\sigma)$ of a graph $G$
and an automorphism $\sigma \in {\rm Aut}(G)$ such that $\sigma$ is not realizable by path flips.
Note that the order of an automorphism $\sigma \in {\rm Aut}(G)$ is the smallest integer $k$ such that $\sigma^k = 1_G$.
Let  $(G, \sigma)$ be a counterexample such that
{\rm (1)} $|V(G)|$ is minimum, and 
{\rm (2)} the order of $\sigma$ is minimum subject to {\rm (1)}.
Let $n$ be the order of $\sigma$.
First, we claim that $n$ is a prime power.
Indeed, if $n$ is not a prime power, there exist two relatively prime integers $r \ge 2$ and $s \ge 2$ with $n=rs$.
Since the order of $\sigma^r$ is $s<n$ and the order of $\sigma^s$ is $r<n$, by the choice of $n$, both $\sigma^r$ and $\sigma^s$ are realizable by path flips.
Since $r$ and $s$ are relatively prime, there exist
two positive integers $x$ and $y$ such that $rx + sy \equiv 1 \pmod{n}$. 
Hence, we have $\sigma = (\sigma^r)^x (\sigma^s)^y$ and so $\sigma$ is also realizable by path flips.

Let $n = p^\alpha$, where $p$ is a prime and $\alpha$ is a positive integer.
Let $C(\sigma)$ denote the cyclic subgroup of ${\rm Aut}(G)$ generated by $\sigma$.
If $\sigma'$ is another generator of  $C(\sigma)$, $\sigma'$ is realizable by path flips if and only if $\sigma$ is  realizable by path flips.
For a vertex $x$ of $G$, let us denote the {\it orbit} of $x$ in $C(\sigma)$ by $C(\sigma) \cdot x$.
Let us choose a pair $(\sigma', x)$, where $\sigma'$ is a generator of $C(\sigma)$ and $x$ is a vertex of $G$ such that 
{\rm (1)} the distance $d_G(x, \sigma'(x))$ is minimum, and 
{\rm (2)} $|C(\sigma) \cdot x|$ is minimum subject to {\rm (1)}.

We redefine $\sigma$ as a chosen element $\sigma'$, and put $d=d_G(x, \sigma(x))$ and $m=|C(\sigma) \cdot x|$.
Note that $m$ is a power of $p$, since $m$ divides $n=p^\alpha$.  
First, we deal with the case, where $d=0$. 
\medskip\\
\underline{Case 1.} $d=0$. 
\medskip\\
In this case, we have $C(\sigma) \cdot x = \{ x \}$ and $m=1$.
Let $G'=G-x$. Then there exists a vertex partition
$V(G') = V_1 \cup V_2 \cup \cdots \cup V_s$ with a positive integer $s$,
where $G[V_i]$ is a connected component of $G'$ for all $1 \le i \le s$.

Note that if two vertices $u$ and $v$ are contained in a common $V_i$ for some $i$, 
then there exists a $uv$-path $P$ not passing through $x$. 
Since $\sigma$ is an automorphism with $\sigma(x)=x$, $\sigma(P)$ is a path not passing through $x$. 
Hence, $\sigma(u)$ and $\sigma(v)$ are also contained in a common $V_j$ for some $j$.
Therefore, $\sigma$ induces a permutation $\tilde{\sigma}$ on $\{ 1, \ldots , s \}$
such that $\sigma(V_i)=V_{\tilde{\sigma}(i)}$ for all $1 \le i \le s$.
Furthermore, if $j$ is contained in $C(\tilde{\sigma}) \cdot i$, $G[V_i]$ and $G[V_j]$ are isomorphic to each other.

Let us denote $\tilde{\sigma}$ as a product of transpositions such that $\tilde{\sigma} = \tilde{\sigma}_t \circ \cdots \circ \tilde{\sigma}_2 \circ \tilde{\sigma}_1$ with $\tilde{\sigma}_k(\ell_k)=m_k$, $\tilde{\sigma}_k(m_k)=\ell_k$ for some indices $\ell_k$ and $m_k$ for all $1 \le k \le t$.
Then we have a sequence of automorphisms $\sigma_1, \sigma_2, \ldots, \sigma_t, \tau$ of $G$ such that $\sigma_k(V_{\ell_k}) = V_{m_k}$, $\sigma_k(V_{m_k}) = V_{\ell_k}$, $\sigma_k^2(v)=v$ for $v \in V_{\ell_k} \cup V_{m_k}$ and $\sigma_k(v)=v$ for $v \not\in V_{\ell_k} \cup V_{m_k}$ for $1 \le k \le t$,
 and $\tau (V_i) = V_i$ for all $1 \le i \le s$.
Since $\tau |_{G[V_i]}$, 
i.e. the restriction of $\tau$ to $G[V_i]$, 
is realizable by path flips for each $i$ by the inductive hypothesis, $\tau$ is also realizable by path flips.
What we need to show is that each $\sigma_k$ is realizable by path flips.
Hence, it suffices to prove the assertion under the condition where $V(G') = V_1 \cup V_2$, $\sigma(V_1)=V_2$, $\sigma(V_2)=V_1$ and $\sigma^2(v)=v$ for all $v \in V(G')$.
In this case, let us take $v_1 \in V_1$ such that $d_G(x, v_1)$ is maximum, and let $v_2 = \sigma(v_1)$.
Then we have  $v_i \in V_i$ for $i=1,2$, $\sigma(v_1) = v_2$ and $\sigma(v_2) = v_1$.

Let $P$ be a path of $G$ from $v_1$ to $v_2$, and set a path $P' = P \setminus \{ v_1, v_2 \}$.
Now, let us flip $P$ with a bijection $\tau$ on $V(G)$, and let us flip $P'$ with a bijection $\tau'$ on $V(G)$ subsequently.
Then we have $(\tau' \circ \tau)(v_i) = v_{3-i} = \sigma(v_i)$ for $1 \le i \le 2$,
 and $(\tau' \circ \tau)(v) = v$ for all $v \not\in \{ v_1, v_2 \}$.
Set $H=G \setminus \{ v_1, v_2 \}$.
Since $\sigma(V(H)) = V(H)$, we have $\sigma |_{H} \in {\rm Aut}(H)$.
By the choice of $v_1$ and $v_2$, $H$ is connected.
Hence, by the inductive hypothesis, $\sigma |_{H}$ is realizable by path flips.
Therefore, $\sigma$, which is $\sigma |_{H} \circ \tau' \circ \tau$,
 is also realizable by path flips, as required.
\medskip\\
\underline{Case 2.} $d \ge 1$. 
\medskip\\
Let us take a shortest path $P=y_0 y_1 \ldots y_{d-1} \sigma(x)$ from $x$ to $\sigma(x)$, where we set $x = y_0$.
Let $Y = V(P) \setminus \{ \sigma(x) \}$.
\medskip\\
{\bf Claim 1.} %
If $0 \le i < j \le d-1$, then $\sigma^s(y_i) \ne \sigma^t(y_j)$ for all integers $s$ and $t$.
\medskip\\
Suppose, for contradiction, that  $\sigma^s(y_i) = \sigma^t(y_{j})$ for some $s$ and $t$.
We have $y_i = \sigma^k(y_j)$, where $k=t-s$.
Since $d(y_i, \sigma^k(y_i)) = d(\sigma^k(y_j), \sigma^k(y_i)) = d(y_j, y_i) < d$, by the choice of $d$, $\sigma^k$ is not a generator of $C(\sigma)$.
Since $C(\sigma)$ is a cyclic group of the order $n = p^\alpha$, we have $k \equiv 0 \pmod{p}$. 
Furthermore, we have $d(y_i, \sigma^{k+1}(y_i))=d(\sigma^k(y_j), \sigma^{k+1}(y_i))$
$= d(y_j, \sigma(y_i))$
$\le d(y_j, \sigma(x)) + d(\sigma(x),\sigma(y_i))$
$= d-j+i < d$.
Therefore, we have $k+1 \equiv 0 \pmod{p}$, a contradiction. \owari 
\medskip\\
{\bf Claim 2.} %
For all $0 \le i \le d-1$, if $s \not\equiv t \pmod{m}$, then $\sigma^s(y_i) \ne \sigma^t(y_i)$.
\medskip\\
Suppose, for contradiction, that $\sigma^s(y_i) = \sigma^t(y_{i})$ for some $s$ and $t$ with $s \not\equiv t \pmod{m}$.
We have $y_i = \sigma^k(y_i)$ with some $k \not\equiv 0 \pmod{m}$.
Since $y_i = \sigma^n(y_i)$ also holds, we have $|C(\sigma) \cdot y_i| \le \gcd(k,n) < m$, because $n$ is a power of $p$ and $k \not\equiv 0 \pmod{m}$.
With the fact $d(y_i, \sigma(y_i)) \le d$, this contradicts the choice of $x$. \owari
\medskip

For $0 \le k \le m-1$, let us define $X_k = \cup_{0 \le i \le n-1{\rm ~and~} i \equiv k \pmod{m}} \sigma^{i}(V(P))$, and $X'_k = X_k - \{ \sigma^k(x) \}$.
Then we have $\sigma(X_k) = X_{k+1}$ for $0 \le k \le m-2$, and $\sigma(X_{m-1}) = X_{0}$.
Furthermore, by Claim 1 and Claim 2,  $X'_k \cap X'_\ell = \emptyset$ for $k \ne \ell$.

Let us define a subgraph $H$ of $G$ such that $V(H) = \cup_{0 \le k \le m-1} X_k$ and
$E(H) = \cup_{0 \le i \le n-1{\rm ~and~} i \equiv k \pmod{m}} \sigma^{i}(E(P))$.
Since $\sigma(V(H))=V(H)$ and $\sigma(E(H))=E(H)$,  $\sigma |_{H}$ is an automorphism of $H$. 
\medskip\\
{\bf Claim 3.} %
$\sigma |_{H}$ is realizable by path flips on $H$.       
\medskip\\
For $0 \le k \le m-1$, let $H_k = H[X_k]$.  
If $m=n$, 
by definition of $X_k$ and $H_k$,
we have $V(H_k) = X_k = \sigma^k(V(P))$, and $E(H_k)=\sigma^k(E(P))$. 
Hence, $H_k$ is simply a path $\sigma^k(P)$ for $0 \le k \le m-1$, and $H$ is a cycle.
Hence, ${\rm Aut}(H)$ is isomorphic to a dihedral group, which is generated by a pair of reflections of cycles.
Since a reflection is realizable by a path flip, the claim is proved.
In the following, we assume that $m < n$.
Let us define a configuration $\tau \in {\cal C}(H)$ such that
$\tau(v) = \sigma(v)$ for $v \in V(H) \setminus  X_{m-1}$ and
$\tau(v) = \sigma^{1-m}(v)$ for $v \in X_{m-1}$.

We claim that $\tau$ is an automorphism of $H$, because for $0 \le k \le n-1$, $\sigma |_{H_{k}}$ is an isomorphism from $H_{k}$ to $H_{k+1}$ and $\sigma^{1-m} |_{H_{m-1}}$ is an isomorphism from $H_{m-1}$ to $H_0$.
Furthermore, by definition, the order of $\tau$ is $m$.
Since $m < n$, by the minimality of $n$, $\tau$ is realizable by path flips.
On the other hand, $\sigma^m |_{H_0}$ is an automorphism of $H_0$.
Since the order of $\sigma^m |_{H_0} = n/m <n$,  by the minimality of $n$, $\sigma^m |_{H_0}$ is realizable by path flips.
Since $\sigma$ is a composition of $\tau$ and  $\sigma^m |_{H_0}$, $\sigma$ is  also realizable by path flips. \owari
\medskip
   
We may assume $V(G) \setminus V(H) \ne \emptyset$.
Choose a vertex $z \in V(G) \setminus V(H)$ such that $d_G(z, V(H))$ is maximum.
Let $Q$ be a shortest path from $z$ to $V(H)$, and let $y \in V(Q) \cap V(H)$ be the end vertex of $Q$.
Then $y$ is contained in $\sigma^i(Y)$ 
for some $i$, where $0 \le i \le n-1$.
Since we have $d_G(\sigma^{-i}(z), V(H))$ $ = d_G(z, \sigma^{i} (V(H)))$ $ = d_G(z, V(H))$,
by replacing $V(Q)$ with $\sigma^{-i}(V(Q))$ if necessary, we may assume $y$ is contained in $Y$ from the beginning.
Let us define a subgraph $F$ of $G$ such that $V(F) = V(H) \cup \cup_{0 \le i \le n-1} \sigma^{i}(V(Q))$ and $E(F) = E(H) \cup \cup_{0 \le i \le n-1} \sigma^{i}(E(Q))$.
Since $\sigma(V(F))=V(F)$ and $\sigma(E(F))=E(F)$, $\sigma |_F$ is an automorphism of $F$.  
\medskip\\
\underline{Case 2.1.} $V(F) \ne V(G)$.
\medskip\\
In this case, by the minimality of $|V(G)|$,  $\sigma |_F$ is realizable by path flips.
Let us define two more subgraphs $F' = F - C(\sigma) \cdot z$ and $G' = G - C(\sigma) \cdot z$.
By the maximality of $d_G(z,V(H))$, both $F'$ and $G'$ are connected, and $\sigma(V(F'))=V(F')$, $\sigma(V(G'))=V(G')$.
Hence,  $\sigma |_{F'}$ and $\sigma |_{G'}$ are automorphisms of $F'$ and $G'$, respectively.
Again by the minimality of $|V(G)|$, $\sigma |_{F'}$ and $\sigma |_{G'}$ are realizable by path flips.
Since $\sigma$ is a composition of $\sigma |_{F}$, $\sigma^{-1} |_{F'}$ and $\sigma |_{G'}$,  $\sigma$ is  realizable by path flips.
\medskip\\
\underline{Case 2.2.} $V(F) = V(G)$.
\medskip\\
For $0 \le k \le m-1$, let us define $W_k = \cup_{0 \le i \le n-1{\rm ~and~}i \equiv k \pmod{m}}  \sigma^i (V(Q))$.    
Note that $W_k$'s are not necessarily disjoint to each other.
Let us define a configuration $\tau \in {\cal C}(F)$ such that
$\tau(v) = \sigma(v)$ for $v \in V(H) \setminus  (X'_{m-1} \cup W_{m-1})$ and
$\tau(v) = \sigma^{1-m}(v)$ for $v \in X'_{m-1} \cup W_{m-1}$.
We need to check that $\tau$ is well-defined.
Suppose that there exists a vertex $v \in V(F)$ such that $v \in  (X'_{m-1} \cup W_{m-1}) \cap  (X'_{k} \cup W_{k})$ for some $k$ with $0 \le k \le m-2$.
Suppose that there exists a vertex $v \in V(F)$ such that $v \in
(X'_{m-1} \cup W_{m-1}) \cap  (X'_{k} \cup W_{k})$ for some $k$ with
$0 \le k \le m-2$.
Then there exists a positive integer $i$ such that $\sigma^{i}(v)=v$
and $i$ is not divisible by $m$.
Let $j$ be a greatest common divisor of $n$ and $i$.
Then we have $\sigma^{j}(v)=v$ and $j$ is not divisible by $m$.
Since both $j$ and $m$ are divisors of $n$, a prime power,
$j$ is a divisor of $m$.
Hence, we have $\sigma^m(v) = v$, which implies $\sigma(v) = \sigma^{1-m}(v)$. 

For $0 \le k \le m-1$, let $F_k = F[X_k \cup W_k]$. 
We claim that $\tau$ is an automorphism of $F$, because $\sigma |_{F_{k}}$ is an isomorphism from $F_{k}$ to $F_{k+1}$ and $\sigma^{1-m} |_{F_{m-1}}$ is an isomorphism from $F_{m-1}$ to $F_0$.
Furthermore, by definition, the order of $\tau$ is $m$.
\medskip\\
\underline{Case 2.2.1.} $m<n$.
\medskip\\
In this case, by the minimality of $n$,  $\tau$ is realizable by path flips.
Furthermore, $\sigma^{m} |_{F_0}$ is an automorphism of $F_0$ and the order of
$\sigma^{m} |_{F_0}$ is $n/m$, which is less than $n$. Hence,
by the minimality of $n$, $\sigma^{m} |_{F_0}$ is realizable by path flips.
Since $\sigma$ is a composition of $\tau$ and $\sigma^{m} |_{F_0}$, $\sigma$ is realizable by path flips.
\medskip\\
\underline{Case 2.2.2.} $m=n$.
\medskip\\
In this case, $H$ is a cycle of order $dn$.
Put $r = dn$.
We relabel the vertices of $F$ as follows:
let us label $V(Q)$ as $Q = w_0 w_1 \ldots w_s$, where $w_0 = z$ and $w_s = y$.
For $0 \le i \le n-1$ and $0 \le j \le s$, let $w_{i,j} = \sigma^i (w_j)$.
Note that $w_{i,j}$ may coincide with $w_{k,j}$ for some $i \ne k$.
We also write $Q_i = w_{i,0} w_{i,1} \ldots w_{i,s}$ for $0 \le i \le n-1$.
Let us label the vertices of $H$, which is a cycle of length $r$, as $H = z_0 z_1 \ldots z_{r}$, where $z_r = z_0$ and $z_{id} = w_{i,s}$ for $0 \le i \le n-1$.

For a positive integer $N$ and for an integer $t$, 
let us define permutations $\pi(N,t)$ on $\{ 0,1,\ldots, N-1 \}$ such that $\pi(N,t)(i) \equiv t-i \pmod{N}$ for $0 \le i \le N-1$.
For an integer $t$, let us define a bijection $\rho_t$ on $V(F)$ satisfying $\rho_t^2 = 1$, as follows:

$\rho_t(w_{i,j})=w_{\pi(n,t)(i),j}$ for $0 \le i \le n-1$ and $0 \le j \le s$, 
and $\rho_t(z_{i})=z_{\pi(r,dt)(i)}$ for $0 \le i \le r-1$.

We need to check that $\rho_t$ is well-defined.
If $w_{i,j} = w_{k,j}$ for some $i,k,j$ with $i \ne k$, we have $\sigma^i(w_j) = \sigma^k(w_j)$.
For any integer $t$, 
since $\pi(n,t)(k)-\pi(n,t)(i) \equiv i-k \pmod{n}$, we have $w_{\pi(n,t)(i),j} = \sigma^{\pi(n,t)(i)}(w_j) = \sigma^{\pi(n,t)(k)}(w_j) = w_{\pi(n,t)(k),j}$, as claimed.

Let $H' = F[\cup_{0 \le i \le n-1} Q_i]$.
Since $\rho_t |_{V(H')}$ is a permutation of $Q_i$ for $0 \le i \le n-1$ and
$\rho_t |_{V(H)}$ is a reflection of $H$, 
$\rho_t$ is an automorphism of $F$.
\medskip\\
{\bf Claim 4.} %
$\rho_t$ is realizable by path flips on $F$.       
\medskip\\
For all $0 \le i \le n-1$ with $i < \pi(n,t)(i)$, let us choose a shortest path $R_i$ from $w_{i,0}$ to $\rho_t(w_{i,0})=w_{\pi(n,t)(i),0}$.
By consecutive path flips of $R_i$ and  $R_i - \{ w_{i,0}, \rho_t(w_{i,0}) \}$, we can exchange $w_{i,0}$ and $\rho_t(w_{i,0})$ for $0 \le i \le n-1$.
In the remaining graph $F'=F- C(\sigma) \cdot w_{0}$, $\rho |_{F'}$ is realizable by path flips by the minimality of $|V(G)|$, as claimed. \owari
\medskip

Since $\sigma$ is a composition of $\rho_1$ and $\rho_0$, by Claim 4, it is realizable by path flips. 
\owari
 
%

%
\end{document}